\documentclass[letterpaper,aps,pre,twocolumn,showcaps,superscriptaddress,showpacs,amsmath,amssymb]{revtex4}
\usepackage{graphicx}

\newcommand{\mean}[1]{\langle #1\rangle}

\renewcommand{\vec}[1]{\mathbf #1}

\newcommand{\lam}{\lambda}
\newcommand{\vhi}{\varphi}
\newcommand{\sig}{\sigma}
\newcommand{\ra}{\rightarrow}
\newcommand{\x}{\vec r}
\newcommand{\nois}{\boldsymbol\xi}
\newcommand{\Dr}{D_\text{r}}
\newcommand{\lm}{\gamma_\text{L}}
\newcommand{\GamL}{\Gamma_\text{L}^\ast}
\newcommand{\GamD}{\Gamma_\text{D}^\ast}
\newcommand{\GamS}{\Gamma_\text{S}^\ast}

\begin{document}

\title{Crystallization in a dense suspension of self-propelled particles}

\author{Julian Bialk\'e, Thomas Speck, Hartmut L\"owen}

\affiliation{Institut f\"ur Theoretische Physik II: Weiche Materie,
  Heinrich-Heine-Universit\"at D\"usseldorf, Universit\"atsstra\ss e 1,
  D-40225 D\"usseldorf, Germany}

\date{\today}

\begin{abstract}
  Using Brownian dynamics computer simulations we show that a two-dimensional
  suspension of self-propelled ("active") colloidal particles crystallizes at
  sufficiently high densities. Compared to the equilibrium freezing of passive
  particles the freezing density is both significantly shifted and depends on
  the structural or dynamical criterion employed. In non-equilibrium the
  transition is accompanied by pronounced structural heterogeneities. This
  leads to a transition region between liquid and solid in which the
  suspension is globally ordered but unordered liquid-like ``bubbles'' still
  persist.
\end{abstract}


\maketitle


Recently, the collective dynamics of self-propelled ("active") particles has
become a topic of intense research~\cite{Vicsek_review,Ramaswamy} resulting in
a wealth of new non-equilibrium phenomena like swarming~\cite{Vicsek,Engel},
clustering~\cite{PeruaniDB2006,Wensink,Yingzi} and active
swirling~\cite{Aranson}. These phenomena have been observed both in dense
bacterial solutions~\cite{Goldstein} and in artificial
microswimmers~\cite{Kagan}. Excellent model systems for self-propelled
particles are colloidal suspensions, where the motility of colloidal particles
can be achieved and steered by magnetic beads acting as artificial
flagella~\cite{DreyfusBRFSB2005}, by catalytic reactions at
Janus-particles~\cite{ErbeZBKL2008}, or by laser-heated metal-capped
particles~\cite{VolpeBVKB2011}.

The purpose of the present Letter is to show that self-motile interacting
colloidal particles in two dimensions still freeze into a crystalline lattice
displaying long-ranged orientational order despite the fact that energy is
injected incessantly. We explore the nature of this non-equilibrium transition
by Brownian dynamics computer simulations of a Yukawa model of self-propelled
particles. We use a minimal model without explicit alignment of particle
orientations. In equilibrium, i.e. in the absence of self-propagation,
freezing and melting of colloidal suspensions is well
understood~\cite{Lowen_review}. But even for passive particles it is known
that freezing is seriously affected and changed under non-equilibrium
conditions, e.g. in a time-oscillatory external force field~\cite{Hoffmann} or
in shear flow~\cite{Wu}. Recently it has also been shown that active matter
can reach steady states with frozen fluctuations~\cite{scha11}.

For self-propelled particles we find that the freezing transition is largely
shifted relative to its equilibrium location. This shift cannot be explained
by a simple scaling using the concept of an effective
temperature~\cite{Wolynes}; quite in contrast to sedimentation profiles of
suspensions~\cite{Palacci} or the long-time diffusion of single propelled
particles~\cite{Golestanian}. Rather, the transition points based on different
criteria for melting and freezing, which agree in equilibrium, diverge. In
particular, the dynamical Lindemann-like melting~\cite{beda85,zahn00} and
freezing criteria~\cite{lowe96,LPS} define a transition region between liquid
and solid characterized by inhomogeneities of the orientational order
parameter.


We study a suspension of $N$ self-propelled particles moving in two dimensions
and immersed in a solvent. Even though the particles are driven we assume that
the solvent remains in equilibrium at the well-defined temperature $T$. The
overdamped motion of the $i$th particle is described through
\begin{equation}
  \label{eq:lang}
  \dot\x_i = -\nabla_iU + f\vec e_i + \nois_i.
\end{equation}
The noise $\nois_i$ models the stochastic interactions with the solvent
molecules. It has zero mean and correlations
$\mean{\nois_i(t)\nois_j^T(t')}=2\delta_{ij}\boldsymbol 1\delta(t-t')$, where
$\boldsymbol 1$ is the identity matrix. Throughout the paper we employ
dimensionless quantities and measure energy in units of $k_\text{B}T$, length
in units of $\rho^{-1/2}$, and time in units of $(\rho D_0)^{-1}$. Here,
$\rho$ is the number density and $D_0$ is the bare diffusion
coefficient. Particles interact pairwise through the repulsive Yukawa
potential
\begin{equation}
  \label{eq:yuk}
  u(r) = \Gamma\frac{e^{-\lam r}}{r}
\end{equation}
with screening length $\lam^{-1}$ and dimensionless coupling parameter
$\Gamma\equiv V_0\sqrt\rho/k_\text{B}T$, where $V_0$ is the bare potential
strength. The total potential energy then becomes
$U=\sum_{i<j}u(|\x_i-\x_j|)$. In addition to the conservative force due to $U$
a constant force $f$ propels every particle in the direction
\begin{equation}
  \label{eq:3}
  \vec e_i \equiv \left(
    \begin{array}{c}
      \cos \vhi_i \\ \sin\vhi_i
    \end{array}\right), \quad
  \mean{\dot\vhi_i(t)\dot\vhi_j(t')} = 2\Dr\delta_{ij}\delta(t-t').
\end{equation}
In the minimal model studied here we assume that these particle orientations
undergo free diffusion without explicit alignment. For spherical particles
with diameter $\sig$ the rotational diffusion coefficient is
$\Dr=3D_0/\sig^2$.


\begin{figure*}[t]
  \centering
  \includegraphics{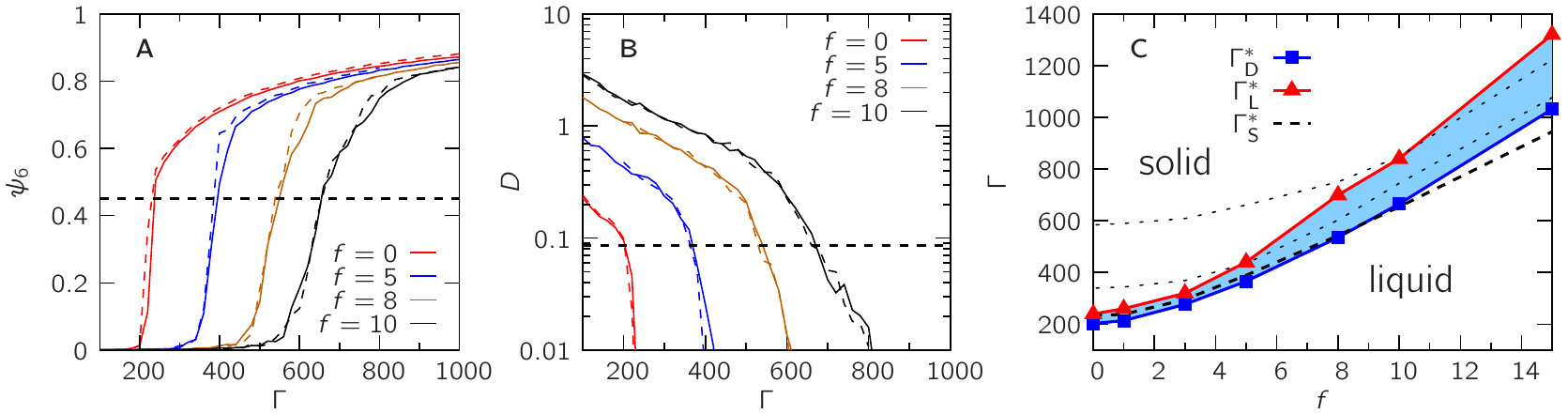}
  \caption{(Color online) Cooling (solid lines) and melting curves (dashed
    lines) for (A)~the orientational order parameter $\psi_6$ and (B)~the
    long-time diffusion coefficient $D$ \textit{vs}. the potential strength
    $\Gamma$ for selected driving forces $f$. The crossings with the dashed
    horizontal lines define the position of the structural transition $\GamS$
    ($\psi_6=0.45$) and the dynamical freezing $\GamD$ ($D=0.086$),
    respectively. (C)~Phase diagram in the $f$-$\Gamma$ plane. The symbols
    mark the numerically estimated dynamical freezing line $\GamD$ and melting
    line $\GamL$ (see main text for definition). The thick dashed line
    indicates the structural transition $\GamS$. Also plotted are the
    $\psi_6=0.67$ and $\psi_6=0.8$ ``iso-structure'' lines along which
    $\psi_6$ is constant.}
  \label{fig:panel}
\end{figure*}


We perform Brownian dynamics simulations for $N=1936$ particles using periodic
boundary conditions. Commensurable box dimensions $L_x/L_y=2/\sqrt 3$ are
chosen such that the suspension can crystallize into the hexagonal crystal
without any defects. We fix the rotational diffusion coefficient to $\Dr=3.5$
and the inverse screening length to $\lambda=3.5$; leaving $\Gamma$ and $f$ as
variable parameters. The time step for updating particle positions is $\Delta
t=10^{-4}$, while particle-particle interactions are cut off after an
inter-particle distance of $7/\lambda=2$.

We simulate cooling and melting runs for forces $0\leqslant f\leqslant
15$. For the cooling runs we use one long trajectory. We start from a random
particle configuration with uniformly distributed orientations. After a
sufficient large relaxation time ($t=25$) we collect data for $500$ time
units. The coupling parameter $\Gamma$ is then increased by 20 and the
protocol of relaxation and recording data is repeated until we reach the
maximal $\Gamma$. The melting runs for each pair $\{\Gamma,f\}$ are
independent starting out of the perfect hexagonal crystal albeit with random
particle orientations. Again, we wait an adequate amount of time before
collecting data for $50$ time units. For both cooling and melting we record
data from 5 independent runs for each $\{\Gamma,f\}$.


We monitor structural changes through the global bond-orientational order
parameter~\cite{stei83}
\begin{equation}
  \label{eq:psi6}
  \psi_6 \equiv 
  \left\langle\left|\frac{1}{N}\sum_{i=1}^N q_6(i)\right|^2\right\rangle,
  \quad
  q_6(i) \equiv \frac{1}{6}\sum_{j\in\mathcal N(i)}
  e^{\text i 6\theta_{ij}},
\end{equation}
where $\mathcal N(i)$ is the set of the six nearest neighbors of the $i$th
particle and $\theta_{ij}$ is the angle between the bond vector pointing from
particle $i$ to $j$ and an arbitrary fixed axis. This order parameter is
practically zero in the disordered phase whereas in a perfect crystal
$\psi_6=1$. In Fig.~\ref{fig:panel}A we plot the $\psi_6$ values averaged over
all runs for both the cooling and the melting protocol. There is a clear
transition between a disordered liquid and an ordered crystalline phase even
for self-propelled particles ($f>0$). However, while the transition is rather
abrupt for $f=0$ the structural ordering is more gradual for higher propelling
forces $f$. In Fig.~\ref{fig:panel}A we do not resolve a possible hexatic
intermediate phase~\cite{bern11}. However, we note that no hysteresis is
observed in agreement with a second-order transition scenario. As a structural
criterion for both the melting and freezing transition we determine $\GamS$
from the condition $\psi_6=0.45$. In particular, for $f=0$ we find
$\GamS\simeq240$, which agrees well with a previous
estimate~\cite{PhysRevE.72.026409}.


Cooling the suspension, a dynamical criterion for freezing is given by the
precipitous drop of the long-time diffusion coefficient
\begin{equation}
  \label{eq:D}
  D \equiv \lim_{t\ra\infty}\frac{1}{4t}\mean{|\Delta\x_i(t)|^2}
\end{equation}
with $\Delta\x_i(t)\equiv\x_i(t)-\x_i(0)$. In Fig.~\ref{fig:panel}B we plot
the diffusion coefficient for different forces. The value $\GamD$ at which the
suspension freezes is estimated from the condition
$D=0.086$~\cite{lowe96}. This gives an upper bound $\Gamma<\GamD$ to the
liquid region, see the phase diagram Fig.~\ref{fig:panel}C. Moreover, for not
too large forces $\GamD\simeq\GamS$ correlates well with the position of the
structural ordering as observed in Fig.~\ref{fig:panel}A. Hence, this
dynamical criterion for freezing based on particle mobility extends only to
weakly driven suspensions of self-propelled particles. Note that at large
forces and small $\Gamma$ the diffusion coefficient $D$ exceeds 1, the
diffusion coefficient of a free passive Brownian particle.

\begin{figure}[t]
  \centering
  \includegraphics{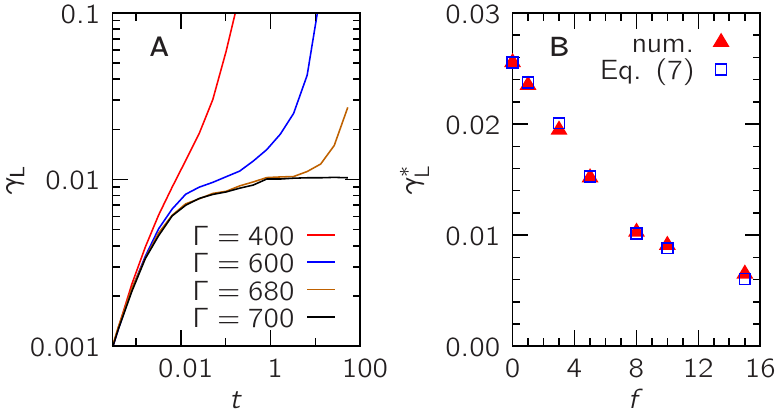}
  \caption{(Color online) (A)~Time-dependence of the Lindemann parameter
    Eq.~\eqref{eq:lm} for $f=8$ below and at the melting point
    $\GamL\simeq700$. (B)~The plateau values of the Lindemann parameter
    $\lm^\ast$ both measured (closed symbols) and from Eq.~\eqref{eq:linde}
    (open symbols) as a function of applied force.}
  \label{fig:linde}
\end{figure}

We next consider a dynamical criterion for melting starting in the solid state
and decreasing $\Gamma$. In one of the first theories for melting Lindemann
conjectured that melting is caused by atom vibrations that start to
interpenetrate~\cite{lind10}. In our case it is natural to consider the
vibrational displacements of particles with respect to their lattice
position. The Lindemann criterion then states that melting commences once the
vibrational displacements reach a certain fraction of the lattice
spacing. However, in two dimensions fluctuations on long wavelengths
eventually destroy long-ranged positional order in the crystal~\cite{merm68}.
The mean-squared displacement, therefore, is not a good measure to distinguish
the liquid from the crystal. Instead, one defines a Lindemann-like
parameter~\cite{beda85,zahn00}
\begin{equation}
  \label{eq:lm}
  \lm(t) \equiv \frac{\mean{|\Delta\x_i(t)-\Delta\x_j(t)|^2}}{2\ell^2}
\end{equation}
from the neighbor-neighbor displacements. Here, $i$ and $j$ denote two
particles that are initially neighbors. The lattice spacing of the hexagonal
crystal is $\ell\equiv2^{1/2}3^{-1/4}\simeq1.075$. In the liquid $\lm(t)$
diverges for long times without a plateau, whereas in the solid one observes a
well defined plateau with Lindemann parameter $\lm$. Hence, we determine the
melting point from the smallest value $\GamL$ for which we still observe a
plateau with value $\lm^\ast$, see Fig.~\ref{fig:linde}A. Above $\Gamma>\GamL$
the suspension is crystalline both with respect to orientational order and a
vanishing diffusion coefficient.

Using a simplified picture to describe the process of melting we assume
particles to move independently close to their lattice position. The
linearized forces then read $-\nabla_i U\approx-k(\x_i-\x_i^0)$ with effective
curvature $k\propto\Gamma$. The initial positions $\x_i(0)=\x_i^0$ correspond
to lattice positions in the hexagonal crystal. A straightforward calculation
of Eq.~\eqref{eq:lm} in the limit $t\ra\infty$ yields
\begin{equation}
  \label{eq:linde}
  2\lm\ell^2 = \frac{4}{k} +
  \frac{2f^2}{k^2-\Dr^2}\left(1-\frac{\Dr}{k}\right).
\end{equation}
In Fig.~\ref{fig:linde}B the plateau value $\lm^\ast$ as a function of force
is plotted together with the prediction $\lm(f,\GamL)$ from
Eq.~\eqref{eq:linde}. Both values show excellent agreement. For the plot we
have fixed the proportionality between $k$ and $\Gamma$ such that the values
for $f=0$ are equal. Moreover, $\lm^\ast(0)\simeq0.026$ agrees well with
previous experiments~\cite{zahn00}.


\begin{figure}[t!]
  \centering
  \includegraphics{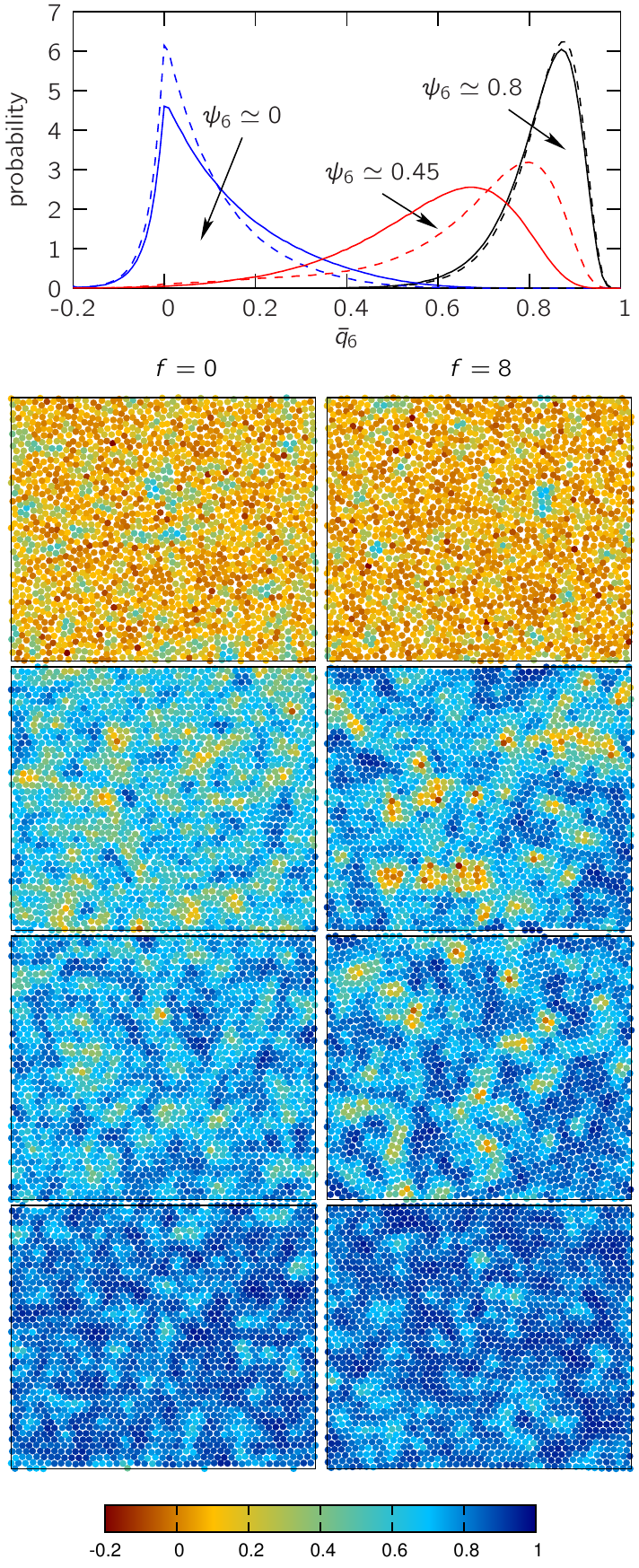}
  \caption{(Color online) Top: Probability distributions for $\bar q_6$ at
    $f=0$ (solid lines) and $f=8$ (dashed lines) for three different global
    $\psi_6$ values. Bottom: Snapshots of particle configurations for both the
    equilibrium ($f=0$, left column) and driven ($f=8$, right column)
    suspension. The rows correspond to constant global $\psi_6$ values: from
    top to bottom $\psi_6\simeq0,0.45,0.67,0.8$,
    cf. Fig.~\ref{fig:panel}C. Particles are colored according to their $\bar
    q_6$ value. While liquid and crystal (top and bottom row) are
    indistinguishable the transition region (middle rows) is marked by
    heterogeneous structure.}
  \label{fig:struct}
\end{figure}

While structural and dynamical criteria agree in equilibrium the phase diagram
Fig.~\ref{fig:panel}C shows that in non-equilibrium there is a transition
region $\GamD<\Gamma<\GamL$ between liquid and crystal, which widens for
larger forces $f$. This region of parameter space is characterized by a high
structural order but non-vanishing long-time diffusion. Moreover, the
dynamical freezing and melting lines do not follow the orientational order but
are shifted to higher $\Gamma$ at higher forces. This implies that at high
propelling speeds structural ordering occurs before dynamical freezing. While
an effective temperature could be defined individually for each criterion, the
resulting values as a function of force clearly do not agree.

To gain further insight we define the order parameter
\begin{equation}
  \label{eq:q}
  \bar q_6(i) \equiv \text{Re}
  \frac{1}{6}\sum_{j\in\mathcal N(i)} q_6(i)q^\ast_6(j)
\end{equation}
per particle in order to effectively describe the local environment of every
particle~\cite{lech08}. The advantage of the neighbor-shell averaging compared
to $|q_6|^2$ is that $\bar q_6$ more sharply distinguishes liquid-like from
ordered regions. In Fig.~\ref{fig:struct} probability distributions of $\bar
q_6$ for $f=0$ and $f=8$ are plotted. For the crystal ($\psi_6\simeq0.8$) no
difference between the driven and the undriven suspension is discernible (see
also the last row of Fig.~\ref{fig:struct}). In the liquid ($\psi_6\simeq0$)
the driven suspension is somewhat less structured compared to
equilibrium. This is caused by the larger effective diffusion due to the
propulsion. A large difference can be seen in the distributions for
suspensions with $\psi_6\simeq0.45$, i.e., in the transition regime. Here the
driven suspension is locally more ordered but with a long tail that extends
down to unordered particles. The spatial distribution of order and disorder
corresponding to this $\psi_6$ for a single snapshot is shown in the second
row of Fig.~\ref{fig:struct}. For $f=8$ the suspension is overall more ordered
but also more heterogeneous, i.e., small, well separated liquid ``bubbles''
remain. Interestingly, the $\psi_6=0.67$ iso-line crosses the melting line
such that for $f=8$ it is within the transition region. Two snapshots for this
case are depicted in the third row of Fig.~\ref{fig:struct}. Due to the
crossing the two forces now also describe two different dynamic regimes: while
diffusion has effectively ceased in equilibrium, some particles still move in
the driven suspension.


In conclusion, we have shown by using Brownian dynamics computer simulations
that self-motile colloidal particles crystallize at sufficiently high
densities. As compared to the equilibrium freezing of passive particles there
is a significant shift in the freezing density and additional large structural
fluctuations appear caused by the self-propulsion. In principle, our
predictions are verifiable in real-space experiments on colloidal model
swimmers on a (quasi) two-dimensional
substrate~\cite{ErbeZBKL2008,VolpeBVKB2011}.

In future work, it would be interesting to generalize our model to one which
embodies an explicit swarming behavior such as a self-propelled rod
model~\cite{Wensink}. Furthermore, since equilibrium freezing is different in
two and three spatial dimensions, it would be very interesting to simulate a
corresponding three-dimensional model. Last but not least the influence of
self-motility of the glass transition has not yet been studied. Since glass
formation competes with crystallization and is typically accompanied with
dynamical heterogeneity~\cite{Richert,Ramos,Chandler10}, self-propulsion may
introduce an internal source of additional fluctuations which can help to form
amorphous structures provided the density is large enough.


We thank H. H. Wensink, G. Volpe, I. Theurkauff, C. Cottin-Bizonne, and
L. Bocquet for helpful discussions. This work was supported by the DFG within
the SFB TR6 (project D3). TS acknowledges financial support by the
Alexander-von-Humboldt foundation.


\end{document}